\documentclass[preprint,prc,aps,showpacs]{revtex4}
\usepackage{epsfig}

\begin{document}

\title{Proton-nucleus scattering and cross section fluctuations at RHIC
 and LHC}

\author{V. Guzey}
\affiliation{Institut f{\"u}r Theoretische Physik II,
Ruhr-Universit{\"a}t Bochum,
  D-44780 Bochum, Germany}
\email[]{vadim.guzey@tp2.ruhr-uni-bochum.de}

\author{M. Strikman}
\affiliation{Department of Physics, Pennsylvania State University,
University Park, PA 16802, USA}
 \email[]{strikman@phys.psu.edu}

\preprint{RUB-TPII-5/2005}

\begin{abstract}

We consider high-energy proton-heavy nucleus scattering within the
framework of the Glauber-Gribov approximation and taking into account
cross section fluctuations. Fixing parameters of the model for cross section
fluctuations by the
available data, we make predictions for the
total, elastic and coherent diffractive dissociation proton-nucleus cross
sections  for the RHIC and LHC energy range.
We predict a strong change of  the $A$-dependence of  
diffraction dissociation from $A^{0.42}$ at RHIC energies
to $A^{0.27}$ at LHC energies.
 Based on the obtained
results, we discuss the approach of the interactions to the
black body (unitarity) limit.
We estimate the electromagnetic contribution
 to coherent $p\,A$ diffraction
and find that it dominates the coherent diffractive
cross section on heavy nuclear targets in the 
RHIC 
and LHC kinematics.

\end{abstract}

\maketitle

\section{Introduction}
\label{sec:intro}

With the advent of the Large Hadron Collider (LHC) one will have
an opportunity to study proton-proton, proton-nucleus and
nucleus-nucleus collisions at the unprecedentedly high energies,
$\sqrt{s}=14$, 9 and 6 GeV per nucleon,
respectively~\cite{HardProbes}. While the main physics drive of
the LHC is the search for Higgs boson, supersymmetry and other
physics beyond the Standard Model, many ideas of the traditional
physics of soft and hard  hadron-hadron interactions can be
tested. In particular, one should be able to address the issue of
blackening of strong interactions at high energies
much better  than this can be done at the RHIC and Tevatron
energies. In this work, the term {\it blackening} means the
approach of a given partial wave its limiting value given by
unitarity of the scattering operator. We refer to this regime 
the black body limit (BBL). Specifically, the TOTEM
collaboration~\cite{TOTEM} at the LHC intends to study the total,
elastic and diffractive dissociation proton-proton cross sections
at the maximal accelerator energy of $\sqrt{s}=14$ GeV with the
aim to test various models, whose predictions depend on the way
the BBL is implemented.

It is commonly believed that phenomena associated with high parton
densities are more pronounced in nuclei than in free nucleons. In
this respect, examining the energy and the atomic mass number $A$
dependence of total, elastic and diffractive dissociation cross
sections in hadron-nucleus scattering, one is expected to see an
enhancement of the effects related to blackening of the proton-proton
interaction.

In this work, we consider total, elastic and diffractive
dissociation proton-nucleus cross sections. As a starting point, 
we use the well-established  Glauber-Gribov multiple scattering
formalism~\cite{Glauber1,Gribov1}, which is known to work with 
a few percent accuracy for total and elastic hadron-nucleus cross
sections.
 While the Glauber method is essentially based on non-relativistic quantum
 mechanics, which takes into account only elastic intermediate states,
its generalization by Gribov within the field-theoretical framework also
includes inelastic (diffractive) intermediate states.
The latter is a manifestation of the increase of the coherence 
length associated with the given process with energy~\cite{Feinberg}.
A convenient way to model this essential feature of high-energy hadron
scattering is by working with eigenstates of the scattering operator and by
introducing cross section 
fluctuations~\cite{GW,Miettinen,Lapidus,Blattel1993,coh_diff,He4}.

The main goal of the present work is to
extend a particular model of cross section fluctuations summarized
in~\cite{Blattel1993} to the RHIC and LHC energies and to make
predictions for the total, elastic and diffractive dissociation
proton-heavy nucleus cross sections and discuss the approach to
the black body regime.

\section{High-energy hadron-nucleus scattering,
 Glauber formalism and cross section fluctuations}
 \label{sec:Glauber}

In order to define and explain   the terms ``black body (disc)
limit'', ``unitarity'',``shadowing'' and
``diffraction'', which we extensively use  in this work, it is
instructive to consider  a simple example of high-energy
scattering on a
 completely absorbing spherical potential with a radius $a$
in non-relativistic quantum mechanics~\cite{Landau3}.
 Making usual
partial wave decomposition, one notices that all partial
scattering amplitudes with the angular orbital moments $l >
l_{{\rm max}}$, where $l_{{\rm max}}=p\, a$ and $p$ is the
projectile momentum, are zero (no scattering). On the other hand,
for the partial scattering amplitudes with $l \leq l_{{\rm max}}$,
scattering is maximal in the sense that there is no transmitted
wave (there is a {\it shadow} formed right behind the target
sphere) and, hence, the scattered wave equals minus the incoming
wave, i.e. the partial scattering amplitudes are $f_l=i/(2 p)$ for
$l \leq l_{{\rm max}}$. Using the optical theorem, one readily
finds the total cross section
\begin{equation}
\sigma_{\rm tot}=2 \pi a^2 \,,
\label{gl1}
\end{equation}
which is twice as large as the geometric cross section of the
target $\pi a^2$. One can separately calculate the elastic cross
section with the result $\sigma_{\rm el}=\pi a^2$  and, hence, the
difference between the total and elastic cross sections, the
inelastic cross section, is $\sigma_{\rm inel}=\pi a^2$.

These classic results can be understood by noticing that the
completely absorbing potential of radius $a$ serves as a {\it
black body} obstacle in the way of the incoming plane wave and
that one deals with {\it diffraction} of the plane wave on a {\it
black disc}. Then in accordance with Babinet's principle of wave
optics, the intensity of the scattered or diffracted light (which
is analogous to $\sigma_{\rm el}$ of our quantum mechanical
exercise) is equal to the intensity of light
 scattered in diffraction on the circular opening of size $a$ in an opaque
 screen, which is proportional to $\pi \, a^2$. At the same time,
 the intensity of the absorbed light, which is analogous to $\sigma_{\rm
 inel}$, is also proportional to $\pi \, a^2$, which means that
  $\sigma_{\rm el}=\sigma_{\rm inel}=\pi a^2$.
The considered example shows that the formation of a shadow behind the
scattering center leads to diffraction. If the scattering
potential is a black body, scattering is maximal and the elastic
cross section (which is, at the same time, the diffractive cross
section) equals half the total cross section. The latter  is twice
as large as the geometric transverse cross section of the target
black disc. A nice discussion of diffraction in high-energy
scattering can be found in~\cite{Goggi}.

In order to show that scattering off the black body is indeed
maximal, we recall the general  condition on the partial
scattering amplitudes, which is a consequence of unitarity of the
scattering operator,
\begin{equation}
Im f_l (\theta)=p \,|f_l(\theta)|^2 +p\,G^{{\rm in}}_l(\theta) \,,
\label{gl2}
\end{equation}
where $G^{{\rm in}}_l$ accounts for inelastic processes; $\theta$
is the scattering angle.
 Solving Eq.~(\ref{gl2}) for $Im f_l (\Theta)$
and choosing the smaller of the two solutions, we obtain
\begin{equation}
Im \, f_l (\Theta)=\frac{1}{2p} \left(1-\sqrt{1-4 p^2 \left(|Re \,
f_l|^2+ G^{in}_l \right)} \right) \,.
 \label{gl3}
\end{equation}
From this equation, one sees that the maximal value of $Im \,
f_l(\Theta)$ is $Im f_l^{{\rm max}} (\theta)=1/(2\,p)$, which
is exactly the value of the scattering amplitude in the
black body scattering problem. One can say that the partial
scattering amplitudes for
 $l \leq l_{{\rm max}}$ {\it saturate}.
 While in the considered simple example blackening of $Im f_l$ leads to
the  energy-independent total cross section, it is not the case
in a more realistic situation. For instance, our analysis will 
demonstrate that the total
proton-nucleus cross section slowly
increases with energy regardless that many partial waves reach
their constant maximal values.

In a number of models, which discuss saturation in hard processes, one
often assumes that the total cross section reaches a fixed maximal value or
that partial scattering amplitudes reach constant values smaller than
the maximal $1/(2 \, p)$, see e.g.~\cite{Golec}.

The choice of the smaller of the two solutions to Eq.~(\ref{gl2}) is a
reflection of the fact that in hadron-hadron scattering, the 
imaginary part of the scattering amplitude is driven by the inelastic
contribution.

Turning to hadron-nucleus scattering, we notice that while the
target nucleus can be better approximated by a completely
absorbing black disk than the target proton, it is still a poor
approximation. A better approach was formulated by
Glauber~\cite{Glauber1}.  The target nucleus is approximated by a
static collection of nucleon scatterers so that the phase of the
elastic scattering amplitude is a sum of the phases accumulated in
each projectile-nucleon scattering. This means that if we express
the elastic hadron-nucleus scattering amplitude $f_A(\vec{q})$ in
terms of the profile function $\Gamma_A(\vec{b})$,
\begin{equation}
f_A(\vec{q})=\frac{i p}{2 \pi} \int d^2 \vec{b}\, e^{i \vec{q} \cdot \vec{b}}\, \Gamma_A(\vec{b}) \,,
\label{gl5}
\end{equation}
then $\Gamma_A(\vec{b})$ can be expressed in terms of the
elementary hadron-nucleon profile functions $\Gamma(\vec{b})$,
\begin{equation}
\Gamma(\vec{b})=\frac{1}{ip \,2 \pi} \int d^2  \vec{q}\, e^{-i
\vec{q} \cdot \vec{b}}\, f(\vec{q}) \,, \label{gl5b}
\end{equation}
 integrated with the nuclear ground state wave function
 $\Psi_A(\vec{r}_1,\vec{r}_2,\dots, \vec{r}_A)$
\begin{equation}
\Gamma_A(\vec{b})=\int d^3\,\vec{r}_1\,d^3\,\vec{r}_2\,\dots d^3\,\vec{r}_A |\Psi_A(\vec{r}_1,\vec{r}_2,\dots, \vec{r}_A)|^2 \left(1-\prod_{i=1}^{i=A}\left(1-\Gamma(\vec{b}-\vec{s}_i) \right)  \right) \,.
\label{gl6}
\end{equation}
Equations~(\ref{gl5})-(\ref{gl6}) assume that at high energies the
 small momentum transfer $\vec{q}$ is perpendicular
 to the direction of the beam, i.e.~it is a two-dimensional vector. The
corresponding conjugated variable is the two-dimensional vector of
the impact parameter $\vec{b}$. In Eq.~(\ref{gl6}), the vectors
$\vec{s}_i$ are the transverse components of the position of the
nucleons $\vec{r}_i$; $f(\vec{q})$ is the hadron-nucleon
scattering amplitude. For sufficiently heavy nuclei ($A > 16$) it
is permissible to neglect the nucleon-nucleon correlations in the
ground state nuclear wave function, which means that each nucleon
moves in the nucleus independently, and to write
\begin{equation}
|\Psi_A(\vec{r}_1,\vec{r}_2,\dots, \vec{r}_A)|^2=\prod_{i=1}^{i=A} \rho_A(\vec{r}_i) \,,
\label{gl7}
\end{equation}
where the nucleon distribution $\rho_A(\vec{r})$ is normalized to
unity. The parameterization of $\rho_A(\vec{r})$ is detailed in
Sect.~\ref{sec:results}.
 Then the nuclear profile function
for a heavy nucleus can be presented in the following compact form
\begin{equation}
\Gamma_A(\vec{b})=1-\exp \left(-A \int d^3 \, \vec{r} \rho_A(\vec{r}) \Gamma(\vec{b}-\vec{s}) \right) \,.
\label{gl8}
\end{equation}
The elementary profile function is readily calculated using the
standard parameterization for the elementary proton-nucleon scattering
 amplitude
\begin{equation}
f(\vec{q})=\frac{i p \,\sigma_{{\rm tot}}(s) \left(1-i \eta\right)}{ 4 \pi}
 e^{-B(s)\, q^2/2} \,,
\label{elem}
\end{equation}
where $\sigma_{{\rm tot}}$ is the energy-dependent total cross section;
$B(s)$ is the slope of the amplitude; 
$\eta=Re \, f(\vec{q}) / Im \, f(\vec{q})$. In our
numerical analysis, we use \cite{He4}
\begin{equation}
B(s)=10.5 +0.5 \,\ln\left(s/s_0\right) \ {\rm GeV}^{-2} \,,
\end{equation}
where $s_0=25$ GeV;
$\eta=\pi/2 \times 0.0808=0.127$.

Evaluating  $\Gamma(\vec{b}-\vec{s})$ using Eq.~(\ref{gl5b}) and
substituting the result in Eq.~(\ref{gl8}), we obtain the Glauber
 approximation expression for
$\Gamma_A(\vec{b})$
\begin{equation}
\Gamma_A(\vec{b})=1-\exp \left(-A/2 \, \sigma_{{\rm tot}}(s)(1-i\eta) T(b) \right) \,,
\label{gl10}
\end{equation}
where
\begin{equation}
T(b)=\int dz \, d^2 \vec{s}  \frac{e^{-(\vec{b}-\vec{s})^2/(2 B(s))}}{2 \pi B(s)}  \rho_A\left(\sqrt{|\vec{s}|^2+z^2} \right) \,.
\end{equation}
In the $B(s) \to 0$ limit, the $T(b)$ function takes a more familiar approximate 
form, $T(b)=\int dz \rho_A(\sqrt{b^2+z^2})$.

It is interesting to point out that the profile function $\Gamma_A(\vec{b})$
 plays the  role of the partial scattering amplitude and the impact parameter
$|\vec{b}|$ plays the role of the orbital momentum $l$. As a
consequence, the unitarity condition is diagonal in  $|\vec{b}|$
and reads (compare to Eq.~(\ref{gl3}))
\begin{equation}
2 Re\, \Gamma_A(\vec{b})=|\Gamma_A(\vec{b})|^2+G^{{\rm in}}(\vec{b}) \,.
\label{gl11}
\end{equation}
The solution to this equation is
\begin{equation}
Re\, \Gamma_A(\vec{b})=\frac{1-\sqrt{1-(1+\eta_A^2) G^{{\rm in}}(\vec{b})}}{(1+\eta_A^2)} \,,
\label{gl12}
\end{equation}
where $\eta_A=Im \, \Gamma_A(\vec{b}) / Re \, \Gamma_A(\vec{b})$.
The maximal value of $Re \, \Gamma_A(\vec{b})$ is unity ($\eta_A$
vanishes in the black disc limit), and, therefore,  the Glauber
approximation expression for $\Gamma_A(\vec{b})$ of
Eq.~(\ref{gl10}) trivially complies with the unitarity constraint
of Eq.~(\ref{gl11}).

The Glauber formalism offers a convenient scheme for the
calculation of various observables measured in the hadron-nucleus
scattering at high-energies such as the total and elastic  cross
sections
\begin{eqnarray}
&& \sigma_{{\rm tot}}^{hA}(s)=2 \int d^2 \, \vec{b} \,Re\,
\Gamma_A(\vec{b})
\,, \nonumber\\
&& \sigma_{{\rm el}}^{hA}(s)=\int d^2 \,\vec{b} \, |\Gamma_A(\vec{b})|^2
\,. \label{gl18}
\end{eqnarray}
It is important to note that while the nuclear profile function
saturates, the scattering cross sections in Eq.~(\ref{gl18}) grow
with energy at large $s$.

The quantum mechanical expressions of the Glauber formalism imply
that coherent diffraction on nuclei consists of only elastic
scattering. This contradicts experiments on diffraction
dissociation, which showed that the incoming particle can
dissociate into states with the same quantum numbers leaving the
target nucleus in its ground state. Therefore, the Glauber
formalism should be extended to accommodate this experimental
fact.

A simple picture of  diffractive dissociation was suggested by
Feinberg and Pomeranchuk~\cite{Feinberg} and elaborated on by
Good and Walker~\cite{GW}. One thinks of the incoming wave as a
coherent superposition of eigenstates of the scattering operator.
Each eigenstate interacts with the target with its own cross
section. Since in general these cross sections (eigenvalues) are
different, the final state contains not only the initial particle
but also other states, which {\it diffracted into existence}.
It is important to note that the formalism of scattering eigenstates
is based on the assumption that one can represent scattering as
superposition of scattering of the components with different interaction
strengths. The use of this assumption and the completeness of the set
 of scattering states allows to obtain compact formulas. 
In perturbative QCD, this
assumption can be justified for $ t \sim 0$ relevant for the scattering off
nuclei, while it is not valid for sufficiently large $t$.

Introducing the probability to interact with a given cross section
$\sigma$, $P(\sigma,s)$, the expressions for the total and elastic
hadron-nucleus cross sections become  (compare to
Eqs.~(\ref{gl18}))
\begin{eqnarray}
&&\sigma_{{\rm tot}}^{hA}(s)=2 \int d \sigma P(\sigma) \int d^2 \, \vec{b} \,Re\, \Gamma_A(\vec{b},\sigma) \,, \nonumber\\
&&\sigma_{{\rm el}}^{hA}(s)=\int d^2 \, \vec{b} \, \left| \int d \sigma
P(\sigma)\Gamma_A(\vec{b},\sigma)\right|^2  \,. \label{csf4}
\end{eqnarray}
In these equations, the profile function
$\Gamma_A(\vec{b},\sigma)$ depends on the eigenvalue $\sigma$
rather than on the total cross section $\sigma_{{\rm
tot}}^{pp}(s)$,
\begin{equation}
\Gamma_A(\vec{b},\sigma)=1-\exp \left(-A/2 \, \sigma(1-i\eta) T(b)
\right) \,.
 \label{scf5}
\end{equation}
Therefore, the cross sections in Eq.~(\ref{csf4}) are sensitive
not only to the first moment of $P(\sigma,s)$, $\langle \sigma
\rangle(s) = \sigma_{{\rm tot}}^{hp}(s)$, but also to higher
moments $\langle \sigma^k \rangle(s)$.

The motivation to introduce cross section fluctuations is the need
for a simple picture of diffractive dissociation.
 The cross section for coherent diffraction dissociation of hadrons
on a nuclear target is found as the difference between the
coherent diffraction and elastic cross sections~\cite{coh_diff},
\begin{equation}
\sigma_{DD}^{hA}(s)=\int d^2 \, \vec{b} \, \left(
  \int d \sigma P(\sigma,s)\left|\Gamma_A(\vec{b},\sigma) \right|^2 -
 \left| \int d \sigma P(\sigma,s)\Gamma_A(\vec{b},\sigma)\right|^2 \right) \,.
\label{scf6}
\end{equation}
Since $\sigma_{DD}^{hA}(s)$ is identically zero if cross section
fluctuations are absent, $\sigma_{DD}^{hA}(s)$ is the most
sensitive observable to cross section fluctuations.

At small impact parameters and large $\sigma$, the nuclear profile
function saturates, $\Gamma_A(\vec{b},\sigma) \approx 1$, and
becomes independent of $\sigma$. This leads to vanishing
$\sigma_{DD}^{hA}(s)$. Therefore, {\it cross section
fluctuations} indicate how close to the {\it black body limit}
regime one is: The proximity to the blackening regime is indicated
by the decreasing size of $\sigma_{DD}^{hA}(s)$.
Phenomenologically this fact can be taken into account by modeling
$P(\sigma)$ which becomes narrower as $\sqrt{s}$ increases and by 
taking into account the increase of $\sigma^{hp}_{{\rm tot}}(s)$
with energy, see
Fig.~\ref{fig:psigma}.

\section{Energy dependence of $P(\sigma,s)$}
\label{sec:Psigma}

The distribution over cross sections $P(\sigma,s)$ has the
following properties~\cite{Blattel1993}:
\begin{eqnarray}
&&\int_0^{\infty} d \sigma P(\sigma,s)=1 \,, \nonumber\\
&& \int_0^{\infty} d \sigma \sigma P(\sigma,s)=\sigma_{{\rm tot}}(s) \,, \nonumber\\
&& \int_0^{\infty} d \sigma \sigma^2 P(\sigma,s)=\langle \sigma^2
\rangle(s) =\sigma_{{\rm tot}}^2(s) \left(1 +\omega_{\sigma}(s)
\right) \,. \label{csf1}
\end{eqnarray}
The first equation is probability conservation; the second
equation requires that $P(\sigma,s)$ reproduces correctly the
total hadron-nucleon cross section; the third equation introduces
$\omega_{\sigma}(s)$ which measures the broadness of cross section
fluctuations around the average value. One can also consider
higher moments of $P(\sigma,s)$.

Equations~(\ref{csf1}) constitute the minimal set of constraints
on $P(\sigma,s)$ and one can successfully model $P(\sigma,s)$
using only these constraints and the behavior of  $P(\sigma,s)$ in
the limiting cases of $\sigma \to 0$
 and $\sigma \to \infty$. The constituent quark counting rules suggest that
$P(\sigma)=\cal{O}(\sigma)$ as $\sigma \to 0$. In addition,
convergence of integrals for the moments of $P(\sigma,s)$ (see
Eqs.~(\ref{csf1})) requires that $P(\sigma,s) \to 0$ faster than
any power of $\sigma$ as $\sigma \to \infty$.

We assume a particular parameterization of
$P(\sigma,s)$~\cite{Blattel1993} and determine free parameters of
the parameterization using Eqs.~(\ref{csf1}) with $\sigma_{{\rm
tot}}(s)$ and $\omega_{\sigma}$ as an input at each energy. In
particular, we use the following form  for the proton
$P(\sigma,s)$,
\begin{equation}
P(\sigma,s)=N(s) \frac{\sigma}{\sigma+\sigma_0(s)}
\exp\left(-\frac{(\sigma / \sigma_0(s)-1)^2}{\Omega^2(s)} \right)
\,, \label{csf2}
\end{equation}
whose parameters at typical energies are summarized in
Table~\ref{table:param}.

It is worth emphasizing that for large $\sigma_{{\rm tot}}^{hN}$ and
for the nuclear observables considered in this paper,
 effects of fluctuations are primarily determined by 
the second moment of $P(\sigma,s)$, i.e.~by the value of the
 dispersion $\omega_{\sigma}$~\cite{coh_diff}. 
This allows us to use a simple form of $P(\sigma,s)$ with
 energy-dependent parameters, 
which still captures the essential features of the distribution over cross sections.

\begin{table}
\begin{tabular}{|c|c|c|c|}
\hline
 $\sqrt{s}$, GeV  &  $\omega_{\sigma}$ & \quad $\Omega(s)$ \quad & \quad $\sigma_0(s)$, mb \quad  \\
\hline
\quad 24 ($n\,D$ data, \cite{Murthy}) \quad & 0.29 & 2.2 & 19  \\
\hline
61 ($p\,D$ data, \cite{Dakhno})   & 0.33 & 3.4 & 16  \\
\hline
546 (UA4, \cite{Bernard}) & \, 0.19 & 0.94 & 48 \\
546 (CDF, \cite{Abe})     & 0.16 & 0.77 & 51 \\
\hline
1,800 (CDF, \cite{Abe}) & 0.15 & 0.72 & 63 \\
\hline
 9,000 (LHC, $p\,A$) & 0.10 & 0.52 & 88 \\
 \hline
14,000 (LHC, $p\,p$ \cite{FELIX}) & \quad 0.065 \quad & 0.39 & 97.5 \\
\hline
\end{tabular}
\caption{Parameters $P(\sigma,s)$ at various typical energies.}
 \label{table:param}
\end{table}

 The total proton-proton cross section $\sigma_{{\rm
tot}}^{pp}(s)$ is calculated using the Regge theory motivated fit
by Donnachie and Landshoff~\cite{DL},
\begin{equation}
\sigma_{{\rm tot}}^{pp}(s)=21.7\,s^{0.0808}+56.08\,s^{-0.4525} \,,
\label{res1}
\end{equation}
which is in a good agreement with the available data. 
Recently more elaborate parameterizations of the total  
proton-proton cross section, which explicitly implement Froissart's 
unitarity bound, were suggested~\cite{Cudell:2001pn,Block:2005pt}.
An inspection shows that all parameterizations predict the values of the total 
proton-proton cross section, which differ by 5-10\% at the Fermilab and LHC
 energies. The nuclear cross sections, which we consider, are 
virtually insensitive to such small differences, primarily due to
 the approximate
 saturation of the nuclear profile function $\Gamma_A(\vec{b},\sigma)$, see the discussion
in the end of Sect.~\ref{sec:Glauber}.
We explicitly checked that all nuclear cross sections presented in our work
change by at most 1.5\%, when instead of the parameterization 
of $\sigma_{{\rm tot}}^{pp}(s)$ of Donnachie and Landshoff~\cite{DL}, we
use the parameterization of~\cite{Cudell:2001pn}. For the parameterization
of~\cite{Block:2005pt}, the change is absolutely negligible.

The parameter $\omega_{\sigma}$ is a key input of our analysis
since it defines the broadness of $P(\sigma,s)$ ($\omega_{\sigma}
\propto \Omega(s)$) and, hence, the magnitude of cross section
fluctuations.
 Information on $\omega_{\sigma}$ can be extracted either from
the inelastic shadowing correction in proton (neutron)-deuterium
total cross section or from proton-proton or proton-antiproton
single diffraction at $t=0$, see the details in \cite{Blattel1993}. For the
lower values of $\sqrt{s}$ and the UA4 point at  $\sqrt{s}=546$
GeV, we used the results of~\cite{Blattel1993}. In particular,
there were used the neutron-deuterium total cross section
data~\cite{Murthy} (with maximal  $\sqrt{s} \approx 24$ GeV), the
analysis of~\cite{Dakhno} of the proton-deuterium data with
maximal $\sqrt{s} \approx 61$ GeV), and
 the proton-antiproton single diffraction data taken by the UA4 experiment
 at the SPS collider at CERN with $\sqrt{s} = 546$ GeV~\cite{Bernard}.

In addition to this, we used the CDF (Fermilab) data on
proton-antiproton single diffraction with $\sqrt{s}= 546$ GeV and
$\sqrt{s}= 1800$ GeV~\cite{Abe}. An
 extrapolation to the LHC
proton-proton energy $\sqrt{s}= 14$ TeV, $\omega_s=0.06-0.07$,
is done using K.~Goulianos fit and is
cited in~\cite{FELIX}. A linear interpolation between the
$\sqrt{s}=1.8$ TeV and $\sqrt{s}=14$ TeV gives an estimate for the
value of  $\omega_s$ at the proton-nucleus LHC energies, $\omega_s
\approx 0.10$.
Note that the uncertainty of the extrapolation of diffraction 
from the Fermilab 
to the LHC energies (the uncertainty in the value of $\omega_s$)
 constitutes the main
uncertainty of our predictions 
for the absolute value of $\sigma_{DD}^{hA}$,
but it 
affects only very weakly
 our predictions for the $A$-dependence
of the diffractive cross section.
This  uncertainty will be   rectified during early runs of  
the LHC by the $p\,p$ experiments which will measure diffraction in 
$p\,p$ scattering  at small $t$.

It is important to note that judging by the the values of
$\omega_{\sigma}$ at  $\sqrt{s}=61$ GeV and $\sqrt{s}=546$ GeV, the
function $\omega_{\sigma}$ reaches its (broad) maximum around the present
RHIC energy range of $\sqrt{s}=200$ GeV.
 In our analysis, we assumed that
$\omega_{\sigma}(\sqrt{s}=200 \ {\rm GeV})=0.3$.

\begin{figure}[h]
\begin{center}
\epsfig{file=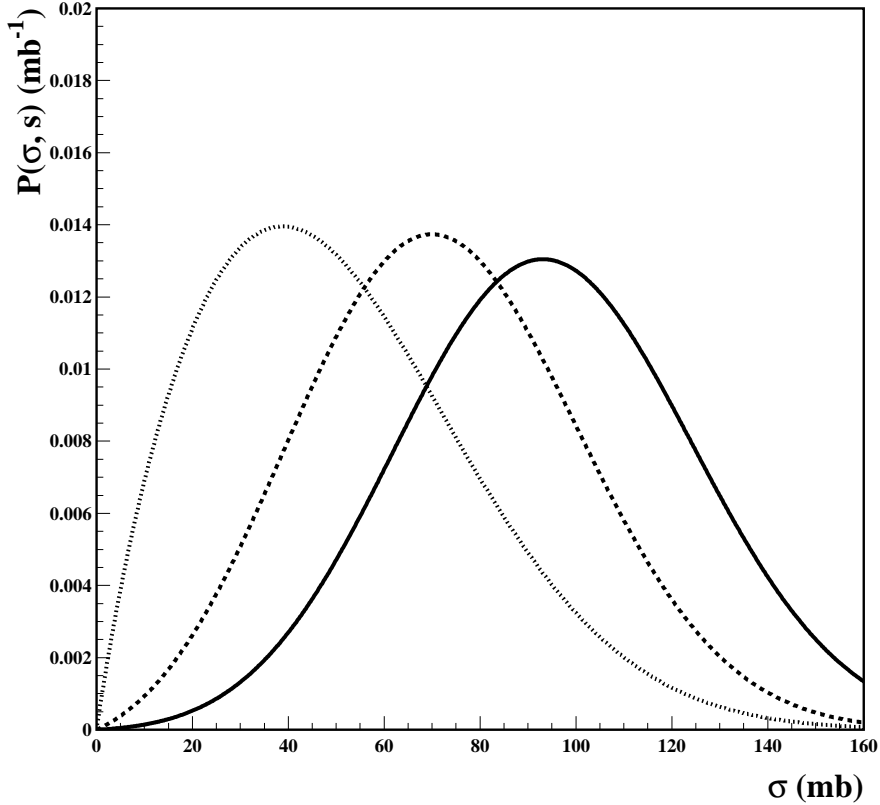,width=12cm,height=12cm} 
\caption{The
cross section distribution $P(\sigma,s)$ at different energies:
the solid curve corresponds to $\sqrt{s}= 9$ TeV (LHC); the dashed
curve corresponds to $\sqrt{s}= 1.8$ TeV (Tevatron); the
dot-dashed curve corresponds to $\sqrt{s}=200$ GeV (RHIC). }
\label{fig:psigma}
\end{center}
\end{figure}

Figure~\ref{fig:psigma} shows the distribution  $P(\sigma,s)$ as a
function of $\sigma$ at three energies considered in
Table~\ref{table:param}: the solid curve corresponds to $\sqrt{s}=
9$ TeV ($p\,A$ collisions at the LHC); the dashed curve
corresponds to $\sqrt{s}= 1.8$ TeV (Tevatron); the dot-dashed
curve corresponds to $\sqrt{s}=200$ GeV (RHIC). As $\sqrt{s}$
increases, the position of the maximum of $P(\sigma,s)$ increases,
which naturally corresponds to the increasing $\sigma_{{\rm
tot}}^{pp}(s)$. 
Although the dispersion $\omega_{\sigma}$  becomes
progressively smaller as the energy increases,
there is no significant change in the width of the distribution 
as measured by the range of values of
$\sigma$, where $P(\sigma, s)>0.5\, {\rm max}P(\sigma,s)$. 
Consequently, even
at the LHC one should
expect significant fluctuations in the number of wounded nucleons
in $p\,A$ scattering at central impact parameters~\cite{Baym1995}.

While the average total cross section increases with energy according to
Eq.~(\ref{res1}), small cross sections can grow with $\sqrt{s}$ much faster. 
For instance, the cross sections corresponding to $P(\sigma,s)=0.002$ in
Fig.~\ref{fig:psigma} increase with energy as
$\sigma \propto s^{0.5-0.75}$.

\section{Results and discussion}
\label{sec:results}

\begin{figure}[t]
\begin{center}
\includegraphics[width=15cm,height=15cm]{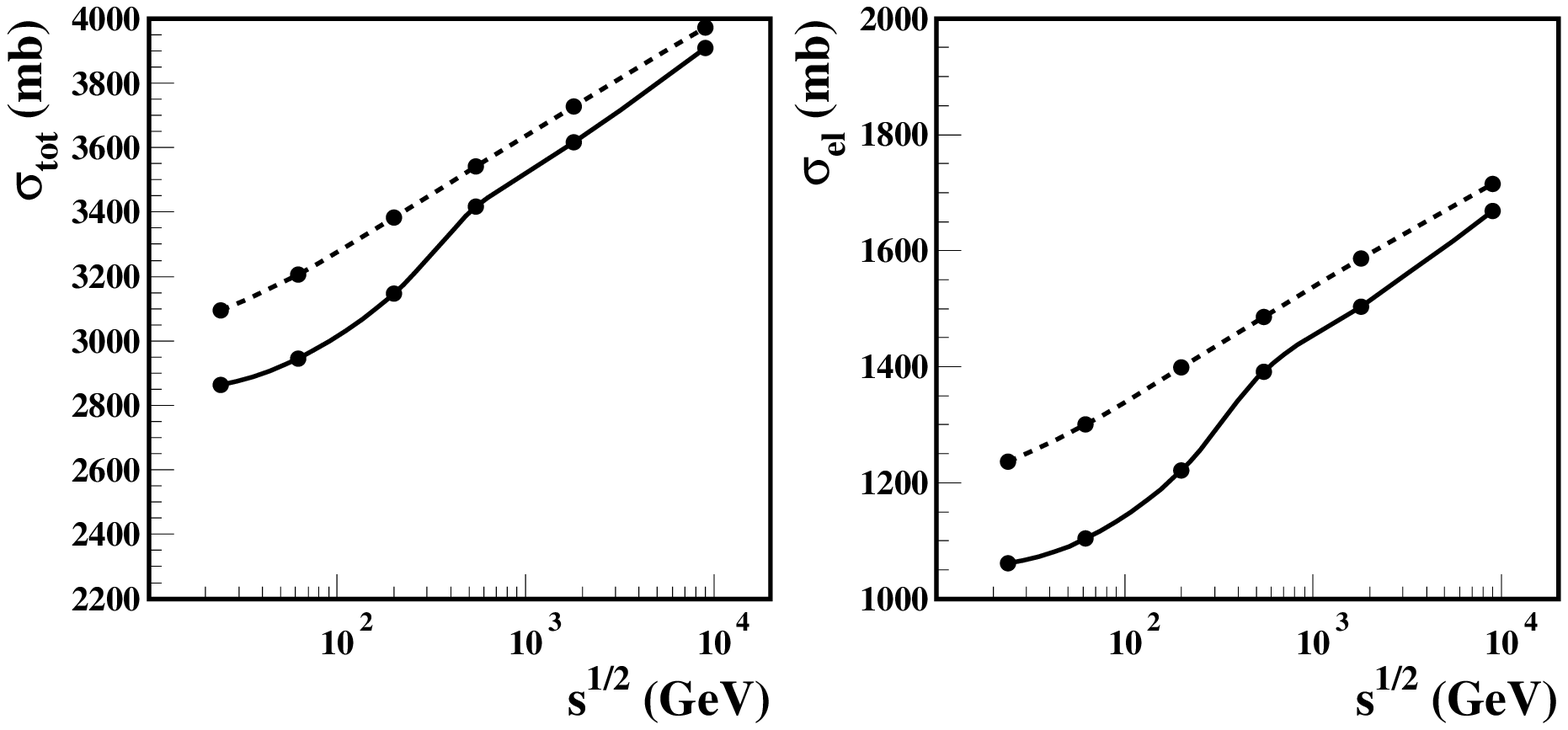}
\vskip -7cm
\includegraphics[width=7.5cm,height=7.5cm]{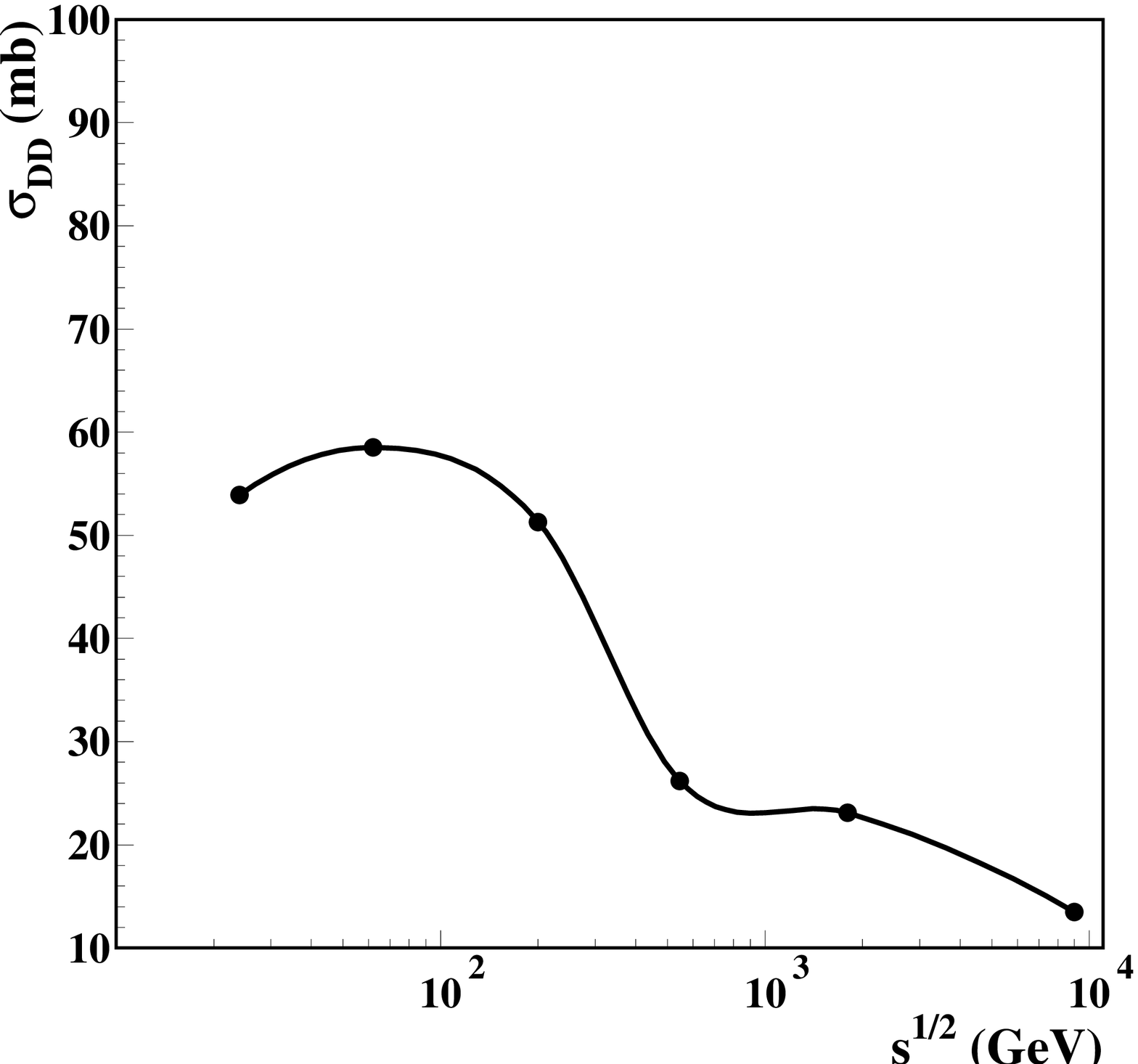}
 \caption{The
proton-Lead total, elastic and diffractive dissociation cross
sections as functions of $\sqrt{s}$. 
The solid curves correspond
to Glauber formalism with cross section fluctuations; the dashed
curves neglect the cross section fluctuations.}
 \label{fig:pb208}
\end{center}
\end{figure}

Using Eqs.~(\ref{gl18}) and (\ref{scf6}), we calculate the total,
elastic and diffractive dissociation cross sections for
proton-$^{208}$Pb scattering as a function of $\sqrt{s}$. The
result is given in Fig.~\ref{fig:pb208}.

\begin{figure}[t]
\begin{center}
\includegraphics[width=15cm,height=15cm]{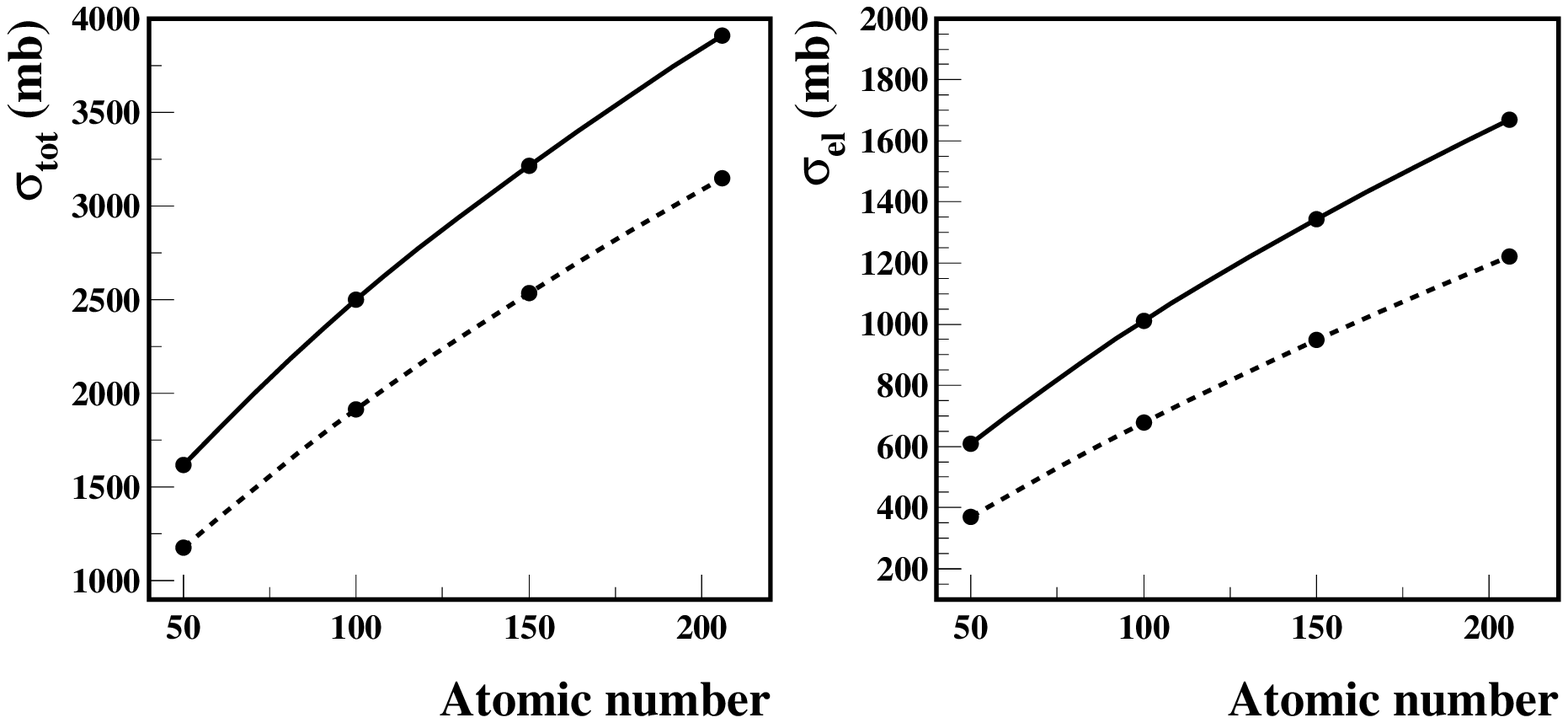}
\vskip -7cm
\includegraphics[width=7.5cm,height=7.5cm]{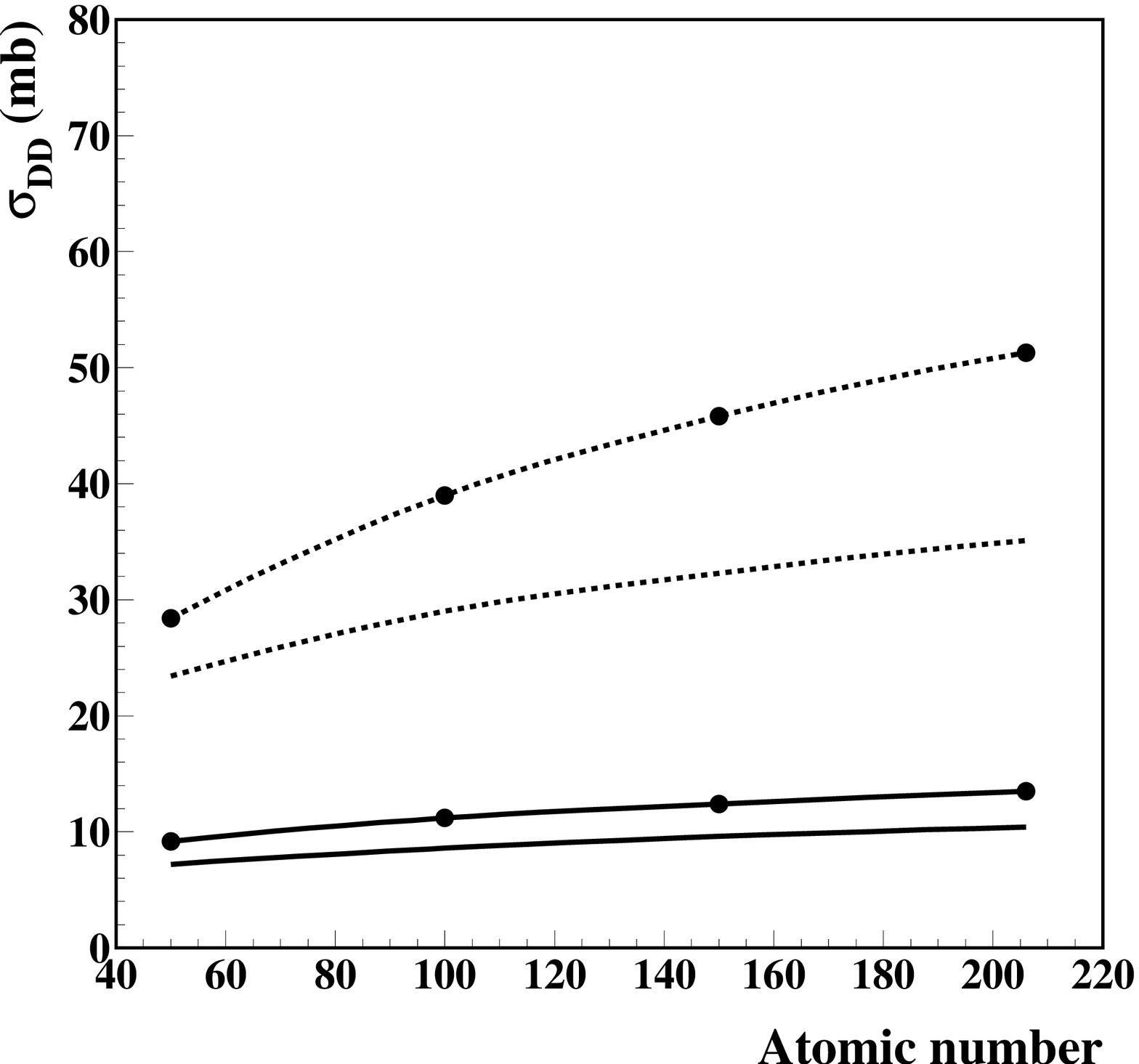}
\caption{The atomic number dependence of the total, elastic and
diffractive dissociation cross sections. The dashed curves
correspond to $\sqrt{s}=200$ GeV and the solid curves correspond
to $\sqrt{s}=9$ TeV.
 The second set of dashed and solid curves, which 
do not go through the points, correspond to the approximate 
calculation of $\sigma_{DD}$ using Eq.~(\ref{eq:approx}).
 } \label{fig:adep}
\end{center}
\end{figure}

In our numerical analysis, we used the following parameterization
of the nucleon distribution $\rho_A(\vec{r})$
\begin{equation}
\rho_A(\vec{r})=\frac{\rho_0}{1+\exp\left((r-c)/a \right)} \,,
\label{res3}
\end{equation}
where $c=R_A-(\pi \, a)^2/(3\, R_A)$ with $R_A=1.145\, A^{1/3}$ fm
and $a=0.545$ fm; the constant $\rho_0$ is chosen to provide the
normalization of $\rho_A(\vec{r})$ to unity.

One sees from Fig.~\ref{fig:pb208} that  cross section
fluctuations decrease the total and elastic cross sections. The
effect is largest in the $\sqrt{s}=100-200$ GeV region. This
can be explained by the increasing role of nuclear shadowing: an
increase of $\omega_{\sigma}$ leads to an increase of the
inelastic shadowing correction, which decreases the total cross
section.

An examination of Fig.~\ref{fig:pb208} shows that, for $\sqrt{s} >
546$ GeV, the total cross section behaves as
\begin{equation}
\sigma_{tot}^{pA}(s) \propto s^{0.045} \,,
\end{equation}
which is slower than the input $\sigma_{tot}^{pp}(s) \propto
s^{0.0808}$.

The diffractive dissociation cross section
 (the lower panel of Fig.~\ref{fig:pb208}) noticeably decreases 
with increasing energies for $\sqrt{s}
> 200$ GeV. 
We would like to stress that the predicted diffractive dissociation 
cross section
primarily depends on the input 
$\omega_{\sigma}$~\cite{coh_diff}
and depends only weakly on the shape of the distribution $P(\sigma,s)$.
Therefore, the diffractive dissociation cross section is a sensitive tool to study the role of cross
 section fluctuations.

We also examined the dependence of the total, elastic and
diffractive dissociation cross sections at $\sqrt{s}=200$ GeV
(RHIC) and $\sqrt{s}=9000$ GeV ($p\,A$ at the LHC) on the atomic
number $A$. The results are summarized in Fig.~\ref{fig:adep},
where the dashed curves correspond to $\sqrt{s}=200$ GeV and the
solid curves correspond to $\sqrt{s}=9$ TeV.

The total cross section behaves with an increasing atomic number as
\begin{eqnarray}
&&\sigma_{tot}^{pA} \propto A^{0.70} \quad {\rm RHIC} \,,
\nonumber\\
&&\sigma_{tot}^{pA} \propto A^{0.62} \quad {\rm LHC} \,.
\end{eqnarray}

The dependence on the atomic number of the  diffractive
dissociation cross section is much slower
\begin{eqnarray}
&&\sigma_{DD}^{pA} \propto A^{0.42} \quad {\rm RHIC} \,,
\nonumber\\
&&\sigma_{DD}^{pA} \propto A^{0.27} \quad {\rm LHC} \,.
\end{eqnarray}
The $\sigma_{DD}^{pA} \propto A^{0.27}$ behavior at the LHC kinematics
 is slower than the $\sigma_{DD}^{pA} \propto A^{0.4}$ result of~\cite{coh_diff}
at much lower energies: cross section fluctuations play a progressively smaller
role as one increases the energy.

It was pointed out in~\cite{coh_diff} that the fluctuations near the average
give the major contribution to $\sigma_{DD}^{hA}$. This point was 
illustrated by Taylor-expanding the integrand in Eq.~(\ref{scf6}) about
$\sigma=\langle \sigma \rangle$ and keeping only first two non-vanishing terms.
The approximate expression for $\sigma_{DD}^{hA}$ reads~\cite{coh_diff}
\begin{equation}
\sigma_{DD}^{hA} \approx \frac{\omega_{\sigma}(s)\sigma_{{\rm tot}}^2(s)}{4}
\int d^2 \vec{b} \left(A T(b)\right)^2 e^{-A \sigma_{{\rm tot}}(s) T(b)}  \,.
\label{eq:approx}
\end{equation}
Note that the effects of $\eta$ are small and can be neglected.
We would like to emphasize that the integral in  Eq.~(\ref{eq:approx})
is a smooth function of $b$, which does not contain a subtraction 
of two large factors, as appears from Eq.~(\ref{scf6}). Therefore,
$\sigma_{DD}^{hA}$ is much more sensitive to the first moments
of $P(\sigma)$, i.e.~to  $\sigma_{{\rm tot}}(s)$ and $\omega_{\sigma}(s)$, rather
than to the details of the shape of $P(\sigma)$.

Calculations of $\sigma_{DD}^{pA}$ using Eq.~(\ref{eq:approx}) are presented in
the lower panel of Fig.~\ref{fig:adep} by the second set of dashed and solid 
curves, which  do not go through the points. For the LHC energy, the approximation
of Eq.~(\ref{eq:approx}) works rather well. For the RHIC energy, the
approximation of Eq.~(\ref{eq:approx}) is good only qualitatively.

\section{Electromagnetic contribution}

Coherent $p\,A$ diffraction, $p+A \to X+A$, has an important electromagnetic 
contribution originating from the ultraperipheral $p\,A$ scattering, when the
 nucleus
acts as a source of quasi-real photons which interact with the proton~\cite{Baur}.
The smallness of the electromagnetic coupling constant
is compensated by nuclear coherence, which gives the enhancement factor $Z^2$,
 where
$Z$ is the nuclear charge. Therefore, the electromagnetic background becomes
important for such heavy nuclei as $^{208}$Pb and constitutes a correction
 for light nuclei down to  $^{40}$Ca.

Since the strong amplitude is imaginary and the electromagnetic one is real, 
the  two contributions do not interfere. Thus,
the cross section of this process is given by convolution of the 
flux of the equivalent photons, $n(\omega)$, with the photon-proton cross
 section,
$\sigma_{\gamma \, p}(\omega)$, see e.g.~\cite{Baur}
\begin{equation}
\sigma_{{\rm e.m.}}^{p A}=\int_{\omega_{{\rm min}}}^{\omega_{{\rm max}}} \frac{d \omega}{\omega}\, n(\omega) \sigma_{{\rm tot}}^{\gamma \, p}(\omega) \,.
\label{bg1}
\end{equation}
In this equation,
\begin{equation}
n(\omega) \approx \frac{2 \,Z^2 \alpha}{\pi} 
\ln \left(\frac{\gamma}{ \omega \, R} \right)\,,
\end{equation}
where $\gamma$ is the Lorentz factor and $R$ is an effective radius of the nucleus;
$\omega_{{\rm max}} \approx \gamma/R$;
$\omega_{{\rm min}}$ determines the minimal photon energy required to excite an 
inelastic final state. Assuming that the lightest inelastic final state in
 the $\gamma \, p$ 
scattering is $\Delta(1232)$, we obtain $\omega_{{\rm min}}=0.3$ GeV.

\begin{figure}[t]
\begin{center}
\epsfig{file=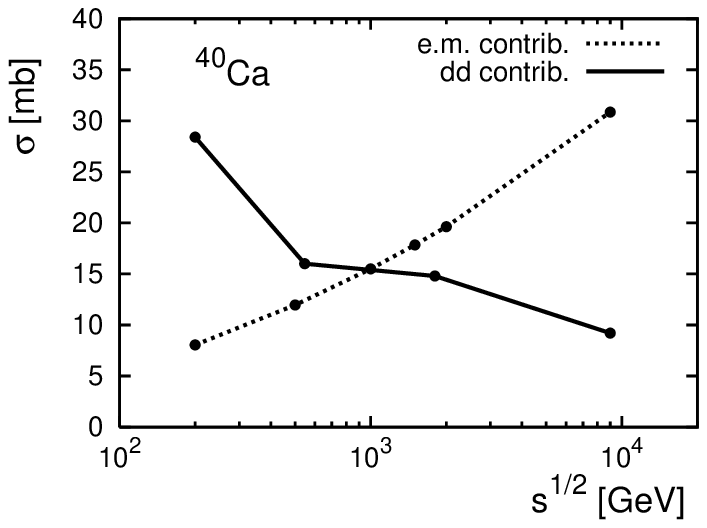,width=8cm,height=8cm} 
\epsfig{file=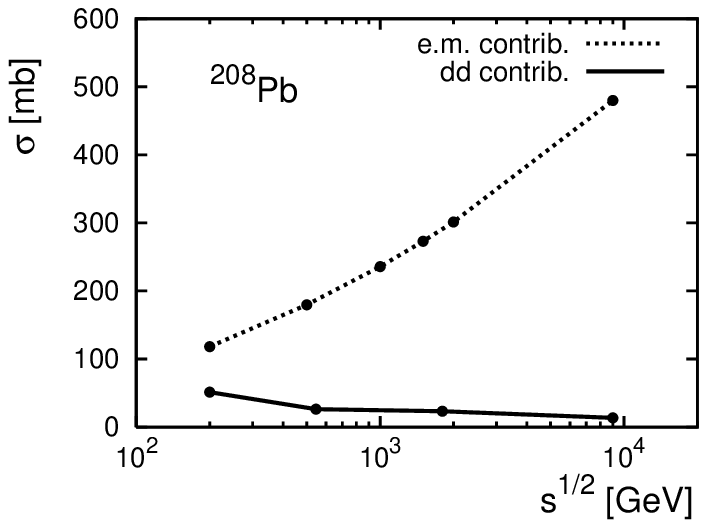,width=8cm,height=8cm}
\vskip 0 cm
\caption{
The electromagnetic contribution evaluated using Eq.~(\ref{bg1})
(dashed curves)
 and coherent diffractive dissociation
cross sections (solid curves) as functions of $\sqrt{s}$ for Pb and Ca.}
\label{fig:em}
\end{center}
\end{figure}

In our numerical analysis of Eq.~(\ref{bg1}), for the Lorentz factor in
the proton rest frame, we used
$\gamma = 2\,\gamma_{\rm L}^A \,\gamma_{\rm L}^p-1$, where 
 $\gamma_{\rm L}^A$ and $\gamma_{\rm L}^p$ are the laboratory frame
Lorentz factors of the nucleus and the proton, respectively. 
This corresponds to $\gamma \approx 2.3 \times 10^{4}$ for RHIC and
$\gamma \approx 4.6 \times 10^{7}$ for the LHC.
The nuclear effective radius was estimated as 
$R=R_A=1.145\, A^{1/3}$ fm, see Eq.~(\ref{res3}).
The real photon-proton cross section was parameterized in the  two-Reggeon 
form of Donnachie and Landshoff~\cite{DL},
\begin{equation}
\sigma_{{\rm tot}}^{\gamma \,p}(s)=0.0677\,s^{0.0808}+0.129\,s^{-0.4525} \,,
\label{bg2}
\end{equation}
where $s=2\, \omega\, m_p+m_p^2$.

The resulting electromagnetic contributions to the coherent diffractive
 cross section are presented in Fig.~\ref{fig:em}
by dashed curves. 
They should be compared to 
the coherent diffractive dissociation cross sections presented by the solid
curves.
The comparison shows that the electromagnetic contribution completely
 dominates coherent $p\,A$ diffraction on Pb-208 at all considered energies.
For the lighter nucleus of Ca-40, the role of the electromagnetic contribution
 becomes progressively important with an increasing energy:
while $\sigma_{{\rm e.m.}}^{p Ca}$ is about 25\% of
 $\sigma^{p Ca}_{DD}$ at the RHIC energy ($\sqrt{s}=200$ GeV),
$\sigma_{{\rm e.m.}}^{p Ca}$ is three times larger than
$\sigma^{p Ca}_{DD}$ in the LHC kinematics ($\sqrt{s}=9000$ GeV).

\section{Conclusions and discussion}

We calculated the total, elastic and diffractive dissociation proton-nucleus
cross sections at high energies using the Glauber-Gribov formalism and
taking into account inelastic intermediate states by means of the
notion of cross section fluctuations.
We extended the model of cross
 section fluctuations of~\cite{Blattel1993}
 to the RHIC and LHC energies
and applied it to the calculation of the cross sections.  
As a consequence of the decrease of cross section fluctuations at the
LHC energy, we observed a significant reduction of
the diffractive dissociation cross section in $p\,A$ coherent diffraction.
This calculation can serve as a benchmark calculation, whose comparison
to the future data can give information on blackening of the proton-proton
interaction.

We found that towards the LHC energies, $\sqrt{s}=9$ TeV, 
$\sigma_{tot}^{pA} \propto s^{0.045}$, which is slower than the input
$\sigma_{tot}^{pp} \propto s^{0.0808}$. Studying the cross sections as
a function of the atomic number $A$, we found that
$\sigma_{tot}^{pA} \propto A^{0.70}$ 
and $\sigma_{DD}^{pA} \propto A^{0.42}$
at  $\sqrt{s}=200$ GeV (RHIC) and that 
$\sigma_{tot}^{pA} \propto A^{0.62}$ 
and $\sigma_{DD}^{pA} \propto A^{0.27}$
at  $\sqrt{s}=9$ TeV (LHC).

Another novel result of the present work is an estimate of  the
 electromagnetic contribution
 to coherent $p\,A$
diffraction coming from  ultraperipheral $p\,A$ scattering.
The electromagnetic smallness of the  background is compensated 
 by nuclear coherence (the enhancement factor is proportional to $Z^2$, where
$Z$ is the nuclear charge) and the Lorentz $\gamma$ factor.
We show that when the nuclear momentum in the laboratory frame is large,
the ultraperipheral e.m.~background
completely dominates  coherent $p\,A$ diffraction on Pb,
see Fig.~\ref{fig:em}.
 One way to reduce the electromagnetic contribution
 is to use lighter nuclei,
 such as for example Ca-40.

\begin{acknowledgments}

The work is supported by the Sofia Kovalevskaya Program of the Alexander
von Humboldt Foundation (Germany) and DOE (USA).

\end{acknowledgments}



\begin{thebibliography}{99}


\bibitem{HardProbes} A. Acardi {\it et al.}, {\it Hard Probes in heavy ion
 collisions at the LHC: PDFs, shadowing and pA collisions},
Subgroup report 3rd Workshop on Hard Probes in Heavy Ion Collisions:
 3rd Plenary Meeting, Geneva, Switzerland, 7-11 Oct 2002, hep-ph/0308248.

\bibitem{TOTEM} TOTEM Collaboration,
{\it Total cross section, elastic scattering and diffraction dissociation
 at the LHC}, CERN/LHCC 97-49, August 1997.

\bibitem{Glauber1} R.J. Glauber, Phys. Rev. \textbf{100}, 242 (1955);\\
V. Franco, Phys. Rev. Lett. {\bf 24}, 1452 (1970).

\bibitem{Gribov1} V.N. Gribov, Zh. Eksp. Teor. Fiz. \textbf{56}, 892 (1969)
[Sov. Phys. JETP \textbf{29}, 483 (1969)]. 

\bibitem{Feinberg} E.L. Feinberg and I.Ia. Pomerancuk, Suppl. Nuovo
Cimento \textbf{III}, 652 (1956).

\bibitem{GW} M.L. Good and W.D. Walker, Phys. Rev. {\bf 120}, 1857 (1960).

\bibitem{Miettinen} H.I. Miettininen and J. Pumplin,
Phys. Rev. D \textbf{18}, 1696 (1978). 

\bibitem{Lapidus} B.Z. Kopeliovich, L.I. Lapidus and A.B. Zamolodchikov,
JETP Lett. \textbf{33}, 595 (1981);
 Pisma Zh. Eksp. Teor. Fiz. \textbf{33}, 612 (1981).

\bibitem{Blattel1993} B. Bl\"attel, G. Baym, L.L. Frankfurt, H. Heiselberg,
 and M. Strikman, Phys. Rev. D {\bf 47}, 2761 (1993).

\bibitem{coh_diff}  L. Frankfurt, G.A. Miller and M. Strikman,
Phys. Rev. Lett. \textbf{71} 2859 (1993).

\bibitem{He4} M. Strikman and V. Guzey, Phys. Rev. C \textbf{52},
1189 (1995).




\bibitem{Landau3} L.D. Landau and E.M. Lifshitz, {\it Course in Theoretical Physics, Vol.3: Non-relativistic quantum mechanics}, 2nd Ed. 1987, Pergamon Press.

\bibitem{Goggi} G. Alberi and G. Goggi, Phys. Rept. \textbf{74}, 1 (1981).

\bibitem{Golec} K. Golec-Biernat and M. Wusthoff,
Phys. Rev. D \textbf{59}, 014017 (1999).



\bibitem{DL} A. Donnachie and P.V. Landshoff, {\it Total cross sections}, Phys. Lett. B {\bf 296} (1992) 227.

\bibitem{Cudell:2001pn}
  J.~R.~Cudell {\it et al.},
  Phys.\ Rev.\ D {\bf 65}, 074024 (2002).

\bibitem{Block:2005pt}
  M.~M.~Block and F.~Halzen,
  Phys.\ Rev.\ D {\bf 72}, 036006 (2005) 
  [Erratum-ibid.\ D {\bf 72},  039902 (2005)].



\bibitem{Murthy} P.V.R. Murthy {\it et al.}, Nucl. Phys. B {\bf 92}, 269 
(1975).

\bibitem{Dakhno} L.G. Dakhno, 
Sov. J. Nucl. Phys. {\bf 37}, 590 (1983) [Yad. Fiz. {\bf 37}, 993 (1983)].

\bibitem{Bernard} UA4 Collaboration, D. Bernard  {\it et al.}, 
 Phys. Lett. B {\bf 186}, 227 (1987).

\bibitem{Abe} CDF Collaboration, F. Abe {\it et al.}, 
Phys. Rev. D {\bf 50}, 5535 (1994).



\bibitem{FELIX} A. Ageev {\it et al.}, {\it A full-acceptance detector at the
LHC (FELIX)}, J. Phys. G: Nucl. Part. Phys. {\bf 28}, R117 (2002).

\bibitem{Baym1995} G. Baym, B. Bl\"attel, L.L. Frankfurt, H. Heiselberg,
and M. Strikman, Phys. Rev. C \textbf{52}, 1604 (1995).


\bibitem{Baur}  G. Baur, K. Hencken, D. Trautmann, S. Sadovsky and
 Y. Kharlov, Phys.Rept. \textbf{364}, 359 (2002) [hep-ph/0112211].


\end{thebibliography}
\end{document}